\documentclass[twocolumn]{emulateapj}

\shorttitle{Warm Halo Cloud Impacts Can Fuel Galactic Nuclei}
\shortauthors{McKernan,Maller \& Ford}

\begin{document}

\title{A New Delivery Route to Galactic Nuclei: Warm Halo Cloud Impacts}

\author{Barry McKernan\altaffilmark{1,2}, Ariyeh Maller\altaffilmark{3}, K.E.Saavik Ford \altaffilmark{1,2}}

\altaffiltext{1}{Department of Science, Borough of Manhattan Community Collge, City University of New York, New York, NY 10007}
\altaffiltext{2}{Department of Astrophysics, American Museum of Natural History, New York, NY 10024}
\altaffiltext{3}{Department of Physics, New York City College of Technology, City University of New York, New York, NY 11201}
\email{bmckernan@amnh.org}

\begin{abstract}
We propose a new mechanism for the delivery of gas to the heart of 
galactic nuclei. We show that warm halo clouds must periodically impact 
galactic centers and potentially deliver a large ($\sim 10^{4-6} 
M_{\odot}$) mass of gas to the galactic nucleus in a singular event. The 
impact of an accreting warm halo 
cloud originating far in the galactic halo can, depending on mixing, produce 
a nuclear starburst of low metallicity stars as well as low luminosity 
accretion onto the central black hole. Based on multiphase cooling around a 
$\Lambda$CDM distribution of halos we calculate the nuclear impact rate, the 
mass 
captured by the central black hole and the fraction of active nuclei for 
impacting cloud masses in the range $10^{4}-10^{6}M_{\odot}$. If there is 
moderate braking during cloud infall, our model predicts an average fraction 
of low luminosity active nuclei consistent with observations.
\end{abstract}

\section{Introduction}
\label{sec:intro}
Activity in galactic nuclei is fuelled by a 
reservoir of low angular momentum gas, but it is unclear how such reservoirs 
build up. Fast outflows from OB stars, supernovae or AGN activity can clear 
out nuclear ISM \citep{b27,b28,b97,b35}. Gas in the 
nucleus can also be quickly consumed by star formation \citep{b36} 
or driven outwards by positive gravitational torques \citep{b37}. If the gas 
in the reservoir originates outside the nucleus, multiple mechanisms 
operating on different distance- and 
time-scales, such as bars feeding nuclear rings \citep{b12}, radiation 
drag \citep{b89} or interactions \citep{b87} are required. From observations,
around $1/3$ of all galactic nuclei in the local Universe exhibit low 
luminosity
 nuclear activity \citep{b66}, so the mechanism promoting nuclear gas build up
is likely to be simple and on-going. 

In this letter we propose a new mechanism for delivery of gas to 
the galactic nucleus. We show that the impact of a warm halo cloud (WHC), 
containing $\sim 10^{4-6} M_{\odot}$, on the central regions of a galaxy will 
fuel nuclear activity. We develop a simplistic model of this phenomenon to 
demonstrate its likely importance. While there remains considerable 
uncertainty in WHC parameters, we show that for plausible input 
parameters, a direct hit by a single WHC on the center of a galaxy will 
supply fuel for star formation and radiatively inefficient accretion 
onto the central black hole. Since WHC bombardment of galaxies must occur, 
some fraction of the low luminosity activity observed in galactic nuclei must 
be due to WHC impacts on galactic nuclei.

\section{Warm Halo Clouds}
\label{sec:hvc}
The assembly of gaseous halos around galaxies naturally produces a multiphase 
medium with warm halo clouds (WHCs) embedded in a low density hot gas halo 
\citep{b96}. WHCs could be the local 
analogue of the high redshift Lyman limit systems \citep{b29}. In the halo
 of our own Galaxy, numerous warm clouds are observed with velocities 
that are inconsistent with Galactic rotation \citep[e.g.][]{b81}. Around 
several other galaxies, HI clouds of $10^{5-8} M_{\odot}$
 have been detected \citep[e.g.][]{b5,b83}. Extended HI 
structures (up to $\sim 10^{10}M_{\odot}$) are observed out to $\sim 100$kpc 
around early-type galaxies \citep{b78}. So, a population of WHCs 
containing $\sim 10^{4-6} M_{\odot}$ per cloud, may be common around most 
galaxies \citep[see e.g.][]{b96}. Around our Galaxy, two basic 
models can account for the properties of observed high velocity clouds 
(HVCs): an accretion model based on WHCs \citep[e.g.][]{b29,b96} or
 a Galactic fountain model \citep{b7}. Accreting WHCs should 
dominate the mass of clouds in the Galactic halo so in this work we shall 
concentrate on the effects of an accreting WHC impact on galactic nuclei. 
Much of the following discussion also applies to 
galactic fountain clouds, although that population 
will have higher metallicity, less mass, smaller radius and lower velocity on 
average, and will be less numerous.

\section{WHC impact with galactic center}
\label{sec:impact}
Large uncertainties exist concerning cloud trajectories around our own Galaxy 
\citep[e.g.][]{b84}. While we might naively expect radial trajectories for 
WHCs, cloud trajectories may become randomized close to a galaxy. Clouds close
 to the disk become tidally disrupted 
\citep{b29,b96}, or deflected by magnetic fields \citep{b1}, or disrupted by 
the Kelvin-Helmholz instability (KHI) \citep{b91} and the fragments dispersed 
about the disk, whereupon the previous cloud trajectory may be irrelevant. 
Since we have no way of predicting actual cloud trajectories, in the
 discussion below we simply assume that clouds impact the disk randomly.

We start by assuming a initial population of $N_{cl}$ clouds on random 
trajectories raining down on a galaxy of radius $R_{gal}$. As long as the 
WHC radius is larger than
 the nuclear region under consideration, the rate of impact ($dN_{cl}/dt$) of 
infalling WHC of radii $r_{cl}$ on the galactic center is 
\begin{equation}
dN_{cl}/dt= 0.25 (r_{cl}/R_{gal})^{2}/\tau
\label{eq:rate}
\end{equation}
 where $\tau$ is the typical cloud infall time and the cloud material is 
assumed to arrive within $0.5r_{cl}$ of the galactic center. We calculate 
the cloud impact rate for multiphase cooling around a 
$\Lambda$CDM distribution of halos based on the model of \citet{b96}. In this
 model the total mass in WHCs is based on the mean free path of the clouds 
and the cloud properties we use are the average of some distribution in the 
halo. First we calculate the average nuclear impact rate, using $R_{gal}=0.15 
R_{cool}$, 
based on conservation of angular momentum \citep{b11}, where $R_{cool}$ is the
cooling radius \citep{b48}. We simplify our calculation by assuming
 that the WHC does not fragment and that infall time is 
$\tau=R_{cool}/V_{max}$ where $V_{max}$ is the maximum circular 
velocity in the halo. The results are shown in 
Fig.~\ref{fig:results}(a), where we plot impacts/Gyr versus $V_{max}$ for 
impacting cloud masses of $10^{4},10^{5},10^{6}M_{\odot}$. Evidently the 
impact rate is relatively flat with $V_{max}$ (or equivalently black hole 
mass). Note that a nuclear impact rate of $\sim 5$/Gyr from 
Fig.~\ref{fig:results}(a) corresponds to a galactic impact rate of 
$\sim 200$/Gyr or a mass inflow rate of $\sim 1M_{\odot}$/yr if the typical 
impactor mass is $\sim 10^{6}M_{\odot}$. This is approximately the low 
metallicity inflow rate required to explain the so-called 
'G-dwarf problem' \citep[e.g.][and references therein]{b16} . 

\section{Delivering material to the Galactic nucleus}
\label{sec:nucleus}
Although observational constraints of cloud impacts are not strong 
\citep{b3}, we can consider the consequences of WHC impact on a galactic 
nucleus, guided by simulations of HVC impacts with the disk 
\citep[see e.g.][and references therein]{b1,b91}. The same basic sources of 
cloud fragmentation in the halo and disk will apply to a WHC falling into a 
galactic bulge. If galactic nuclei have relatively strong magnetic field 
strengths in general \citep[e.g.][]{b94}, the magnetic fields will act as a 
strong brake on infalling WHCs and the cloud may fragment.

Instabilities in the infalling cloud have growth times of the order of the 
timescale on 
which shocks cross the cloud. This cloud crushing timescale, $t_{cc}$, is 
given by \citep{b8}
\begin{equation}
t_{cc}=\left(\frac{n_{cl}}{n_{m}}\right)^{\frac{1}{2}} \frac{r_{cl}}{v_{cl}}
\label{eq:crush}
\end{equation}
where $n_{cl}$ and $n_{m}$ are the densities of the WHC and the 
surrounding medium respectively and 
$v_{cl}$ is the relative velocity of the WHC and the surrounding medium. If the
 WHC is accreting from a very large distance, $t_{cc}$ can be less than the 
infall time ($\sim$ Gyrs) and the cloud can 
fragment in the halo. Indeed many of the HVCs presently observed may be 
fragments of initially much larger accreting WHCs. Galactic haloes are 
hot and diffuse \citep[e.g.][and references therein]{b40,b95} and the 
disruption of
 a WHC in the halo could create an infalling stringy association of clouds 
like the HVC A and C complexes. If the trajectory of such a 
stringy cloud complex intercepted the galactic nucleus then most of the WHC 
mass could be delivered to the region around the central supermassive black 
hole. 

Sufficiently dense and fast clouds falling through the halo and then 
the bulge ISM will first experience deceleration and compression (pancaking), 
followed by an 
expansion of the shocked cloud downstream, then lateral expansion and finally 
cloud destruction when instabilities and differential forces fragment the 
cloud. This entire 
process takes place over a few $t_{cc}$ \citep[see e.g.][]{b8,b1}. 
Compression will increase the WHC density as the cloud pancakes 
\citep{b1,b91}. Small, fast, dense 
fragments of cloud can also sweep up large amounts of gas and be shocked 
\citep{b4}. The transit time for a WHC through a bulge of radius $R_{b}$ to 
the galactic center is $\sim R_{b}/\eta v_{cl}$ where $\eta=[0,1]$ is a 
coefficient incorporating slowdown due to drag and magnetic fields. For a 
galaxy the size of our own($R_{b} \sim 3$kpc), a cloud with 
$v_{cl}\sim 100$ km $\rm{s}^{-1}$ will take $\sim 30/\eta$Myr to 
cross the bulge and reach the center. 

But will the cloud survive this long? Outside of the inner $\sim$0.5kpc, the 
bulge ISM in our own galaxy has very 
low density ($<10^{-3} \rm{cm}^{-3}$) \citet{b9}. At these densities, the 
cloud crushing timescale is $>100$Myrs, far larger than the transit time 
through this part of the bulge. Within $0.5$kpc of the Galactic center the 
bulge ISM density increases to $\sim 0.01\rm{cm}^{-3}$ \citep{b9}. Here the 
crushing timescale is $\sim 30/\eta$Myr, short enough that the WHC will 
fragment in this region. So, WHCs could survive until they 
reach the central regions, whereupon fragmentation is likely. 

\section{Gravitational capture by the central black hole}
\label{sec:next}
Evidently a WHC cloud impact could deliver a large 
quantity of moderate density gas to a galactic nucleus in a single cloud 
infall. But what happens once fragments of a cloud of mass $M_{cl}\sim 
10^{4-6}M_{\odot}$ arrives in the central few tens of pc of a galaxy? 
Low angular momentum fragments of the WHC will be captured by the central 
black hole out to a radius ($R_{esc}$) where their velocity is below the 
escape speed. Thus the mass captured is
\begin{equation}
M_{capt}=M_{cl} \frac{R_{esc}^{2}}{r_{cl}^{2}}=M_{cl} \frac{(2GM_{BH})^{2}}{(\eta v_{cl})^{4}r_{cl}^{2}}.
\label{eq:mcapt}
\end{equation}

We calculate the average $M_{capt}$ for clouds of mass $10^{4-6}M_{\odot}$ 
impacting galactic nuclei. We 
estimate black hole mass using the $M_{BH}-\sigma$ relation from \citet{b15} 
and we assume that the bulge velocity dispersion and maximum circular 
velocity are related as in \citet{b49} so that
\begin{equation}
\rm{log}(\rm{M_{BH}})=5.05\rm{log}(\rm{V_{max}}) -4.41
\label{eq:sigma}
\end{equation}
which should be valid over the range $\sigma_b \sim 70-350$km $\rm{s}^{-1}$. 
We allowed different values of braking ($\eta$), ranging from none ($\eta=1$)
 to strong ($\eta=0.1$) which are shown in Fig.~\ref{fig:results}(b). Clearly 
the braking parameter
 determines when the average mass captured reaches the total cloud mass. Our 
model predicts that the Eddington ratio will be greatest near a critical 
value of
 $V_{max}$, where the average mass captured reaches the total cloud mass.  
Evidently a nuclear strike on a black hole with mass $< 10^{7}M_{\odot}$ 
yields much less mass available for accretion onto the black hole, even for 
large braking ($\eta=0.1$). However, a much larger fraction of 
$M_{cl}$ could be captured if the infalling cloud fragments into a stringy 
cloud complex in the galactic halo, effectively reducing $r_{cl}$ for a given
 $V_{max}$. Cloud material not captured by the black hole will instead be 
gravitationally captured by the bulge. In this case, the material will 
either persist as part of the bulge ISM or will go down the route of star 
formation (with some unknown level of mixing). This is consistent with 
observations of nuclear activity due to $H_{\rm II}$ regions predominating in 
galaxies with smaller bulges and therefore smaller mass central black holes 
\citep{b66}. 

\section{Activity in galactic nuclei}
\label{sec:activity}
What sort of nuclear activity will result from a WHC impact? Once moderately 
dense, low angular momentum fragments 
of the WHC arrive in the galactic nucleus, star formation is likely, although
dependent on poorly 
constrained variables such as cooling rate and mixing with nuclear gas. The 
arrival of a large quantity of infalling gas may also 
initiate activity via shocks or gradual acceleration of GMCs \citep{b85}. In 
our model, shocked nuclear GMCs or the highest 
density, shocked fragments resulting from a WHC impact could generate a 
nuclear OB association within 
$\sim 10$Myr after impact \citep{b10}. A direct hit by a WHC fragment on the 
central black hole may 
result in a phase of low luminosity Bondi accretion. 

\begin{figure}
\epsscale{1.0}
\plotone{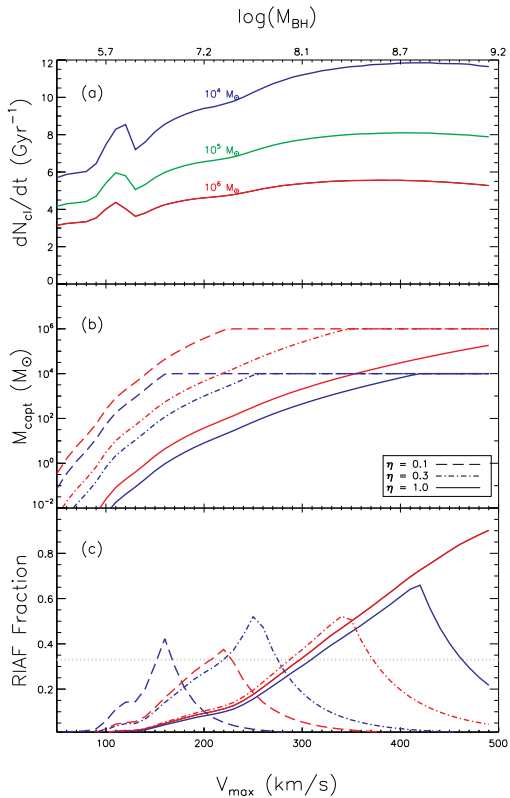}
\caption{Plots of: (a) Average WHC impact rate (number/Gyr) on galactic 
nuclei versus maximum halo circular velocity (km/s) for impactor cloud masses
 of 
$10^{4}$(red),$10^{5}$ (green),$10^{6}M_{\odot}$ (blue). The corresponding 
black hole mass is shown on the top axis. (b) 
Average mass of $10^{4}M_{\odot}$ (red) and $10^{6}M_{odot}$ (blue) impacting
 clouds 
captured by the central black hole versus maximum halo circular velocity 
(km/s)  for a range of braking ($\eta=0.1$,dashed lines;$\eta=0.3$,dot-dash 
lines, $\eta=1.0$, solid lines). (c) Average fraction of 
activity in nuclei due to radiatively inefficient accretion versus  maximum 
halo circular velocity (km/s) for $10^{4}M_{\odot}$ (red) and 
$10^{6}M_{\odot}$ (blue) impacting clouds for a range of braking 
($\eta=0.1-1$) and a range of 
cloud densities (see text for details). The horizontal dotted line 
corresponds to an activity fraction of $1/3$ \citep{b66}. 
\label{fig:results}}
\end{figure}

\citet{b50} find two distributions of Eddington ratios among active 
galactic nuclei in the local Universe. Their first, a lognormal distribution,
 centered around a few percent of Eddington, is associated with nuclear star 
formation, dominates among black holes $<10^{8}M_{\odot}$ and depends 
critically on feedback from luminous accretion. The 
second distribution is a powerlaw which occurs in galaxy bulges where there 
is little or no ongoing star formation. We will only model the second mode 
since the first mode has a complex association with star formation, stellar 
evolution and AGN feedback. Although the distribution of dM/dt in this second
 accretion mode is
 a powerlaw, we will use the average value from our $\Lambda$CDM calculations 
to get an average activity 
fraction. We calculate the average fraction of galaxies of a given $V_{max}$ 
that are active 
for impactors of mass $10^{4-6}M_{\odot}$ and a range of braking and cloud 
densities. The lifetime of Bondi accretion is $\tau_{B}=M_{capt}/(dM/dt)$ 
where 
\begin{equation}
dM/dt \approx 10^{-3} n M_{8}^{2}(\eta v_{200})^{-3} M_{\odot}/yr.
\label{eq:bondi}
\end{equation}
Here $n$ is the average density of the accreting gas, $M_{8}$ is the black 
hole mass in units of $10^{8}M_{\odot}$ and $v_{200}$ is an average 
combination of gas 
speed and gas sound speed in units of 200 km/s \citep{b66}. The average gas 
density 
around the black hole is fairly unconstrained in our model and clearly there
 is trade off
between more massive clouds that take longer to consume and higher gas 
densities which are consumed more quickly. Nevertheless, in this model the 
lowest average  dM/dt is $\sim 10^{-5}M_{\odot}$/yr around a 
$10^{8}M_{\odot}$ black hole, assuming $\eta,v_{200} \sim 1$ and 
$n\sim 0.01 \rm{cm}^{-3}$. In Fig.~\ref{fig:results}(c) we show 
some average values of braking and gas density that yield reasonable models 
where on average 1/3 of galactic nuclei are active. For $10^{4}M_{\odot}$ 
clouds this
requires $n=0.004\rm{cm}^{-3}$, while for $10^{6}M_{\odot}$ clouds with some 
braking this requires $n=0.03\rm{cm}^{-3}$, and with no braking, 
$n=0.01\rm{cm}^{-3}$. Recall we are modelling the powerlaw accretion mode of
 \citet{b50}, which is significant among black holes of mass 
$>10^{8}M_{\odot}$. Therefore, from Fig.~\ref{fig:results}(b),(c), our model 
requires $\eta>0.5,0.3$ for impactor clouds of masses $10^{4}M_{\odot}$ and 
$10^{6}M_{\odot}$ respectively. Note that in all cases the plot shows a 
characteristic shape where the average activity first increases as the amount
 of the cloud captured 
increases and then decreases as the accretion rate increases for the same 
amount of captured cloud. Our model predicts that 
radiatively inefficient Bondi accretion should be more commonly 
detected around larger mass black holes, which is consistent with 
observations of LINER activity in the local Universe 
\citep[e.g.][]{b28,b66,b50}. Future 
observations of low luminosity activity in galactic nuclei as a function of 
$V_{max}$ will be able to rule out certain parameters. Of course if $\eta$ or
 $n$ depend on $V_{max}$ then the shapes of the curves in 
Fig.~\ref{fig:results}(c) might differ. MHD simulations of cloud 
fragmentation in the nucleus are required to establish whether the 
distribution of densities and braking can in fact generate a powerlaw 
distribution of dM/dt \citep{b50} and we intend to do this in future work. 
Note that our model of Bondi accretion is 
unlikely to apply for $M_{capt}<M_{cl}$ since the dominant activity in that 
case would be nuclear star formation, which is consistent with observations 
of smaller mass central black holes \citep{b66}. \citet{b53} argue that 
luminous accretion can occur in the aftermath of a nuclear starburst. This 
may yield the lognormal distribution of Eddington ratio observed in a 
fraction of 
galactic nuclei \citep{b50}, but establishing this is far beyond the scope of
 the present work.

We expect the number of WHCs to increase with redshift as the gas 
accretion rate onto galaxies increases. We naively expect the number of WHC
 clouds and therefore the average cloud impact rate to scale with star 
formation rate or $\propto (1+z)^{3.1}$ from $z=0$ to $z=1$ \citep{b86}, 
corresponding to a rate of $\sim 1/30$Myr for Milky-Way sized galaxies at 
$z \sim 1$. On these timescales, luminous nuclear activity could result if 
radiatively efficient accretion occurs onto the central supermassive black 
hole. A strongly decreasing Eddington ratio over cosmic time 
\citep[e.g.][]{b87,b65,b64} could naturally produce a high duty cycle at 
$z\sim 0$, consistent with Fig.~\ref{fig:results}(c). 
At even earlier epochs ($z>2.5$), the bulk of the growth of supermassive
 black holes occurs via mergers in massive protogalactic halos ($\sim 
10^{11-13} M_{\odot}$) \citep{b88}. At that time, gas accretion is primarily 
on random trajectories, so nuclear fueling by WHC impact would be at its 
maximum. A natural bridge can then be made to the time before hot halos are 
established \citep{b93,b74}, when gas accretion is all in the form of 
infalling gas. In the present epoch, our model predicts that 
clustering of activity should be random for similar sized galaxies 
(comparable $V_{max}$), since cloud impacts are random. This is consistent 
with observations of clustering in LINERs \citet{b77}.

\section{Conclusions}
\label{sec:conclusions}
In this letter we introduce a new mechanism that delivers large quantities of 
gas to galactic nuclei on astrophysically interesting timescales. We show that
 a single warm halo cloud (WHC) impact on a galactic bulge could potentially 
deliver a large mass ($\sim 10^{4-6} M_{\odot}$) of gas to the central regions
 of a galactic nucleus, in a singular event. Although there are considerable 
uncertainties in the parameters of 
warm halo clouds, some representative numbers suggest that the impacts occur 
on astrophysically interesting timescales and at a rate that must account for
some or even all of the low luminosity activity observed in galactic nuclei 
in the local 
Universe. Based on analytic $\Lambda$CDM calculations of cooling halos, 
our model predicts 
an impact rate relatively independent of galaxy mass. Our 
model also predicts that larger mass black holes ($>10^{7}M_{\odot}$) can 
capture significant fractions of the impacting cloud mass at some critical 
value depending on the nuclear braking ($\eta$). At this critical value, our 
model predicts the highest Eddington ratios. Below this critical mass, our 
model predicts that most of the cloud mass will not accrete onto the black 
hole. Instead, this material will mix with the ISM in the nuclear bulge or
 induce star formation, which may lead to delayed episodes of high Eddington 
ratio accretion. Finally, our model 
predicts that, for a reasonable range of cloud masses, densities and braking,
 the fraction of supermassive black holes accreting at very low Eddington 
ratios in the local Universe could be around $\sim 1/3$, which agrees with 
observations \citep{b66}.

\section*{Acknowledgements}
BM \& KESF gratefully acknowledge the 
support of the Department of Astrophysics of the American Museum of Natural 
History, PSC grant PSC-CUNY-40-397 and CUNY grant CCRI-06-22. AM acknowledges 
support from an ROA supplement to NSF award AST-0904059. We acknowledge 
useful discussions with Mordecai-Mark Mac Low and a very helpful report from 
the anonymous referee.


\label{lastpage}

\end{document}